\documentclass[showpacs,pre,amssymb,amsmath,preprint]{revtex4}

\usepackage{psfrag}
\usepackage{amsmath}
\usepackage{amssymb}
\usepackage{textcomp}		% pour le micrometre de la table 4
\usepackage{array}		% pour la figure 10 !
\usepackage{multirow}		% pour le tableau 1
\usepackage{rotating}		% pour le texte vertical dans le tableau
\usepackage{floatflt}
\usepackage{textcomp}		% pour le micrometre de la table 4
\usepackage{color}
\usepackage{tikz}
\usetikzlibrary{decorations.pathreplacing,angles,quotes}
\usetikzlibrary{shapes}
\usepackage{physics}
\usepackage{enumitem}  		%enumerate style romain i ii iii 
\usepackage{setspace}
\usepackage{bm}

\usetikzlibrary{positioning,decorations.pathmorphing}

\numberwithin{equation}{section}

\newcommand{\be}{\begin{equation}}
\newcommand{\ee}{\end{equation}}
\newcommand{\bea}{\begin{eqnarray}}
\newcommand{\eea}{\end{eqnarray}}
\def \la{\label}

\def\({\left (}
\def\){\right )}
\def\]{\right]}
\def\[{\left[}
\def\<{\left <}
\def\>{\right>}

\usepackage{xcolor}
  
\begin{document}

\title{The dilute interacting Bose gas : comparisons between Feynman diagrammatic expansion and the hierarchy equations for the imaginary time Green functions} 

\author{Victor Dansage$^{1,2}$, Vincent Ballenegger$^1$, and Angel Alastuey$^2$
       \\[2mm]
       $^1${\small Institut UTINAM, UMR 6213 CNRS, Université de Franche-Comté,}\\[-1mm]
        {\small 16 Route de Gray, 25030 Besan\c con, France}\\[-1mm]
              $^2${\small Laboratoire de Physique, ENS Lyon, UMR CNRS 5672} \\[-1mm] 
							{\small 46 all\'ee d'Italie, 69364 Lyon Cedex 07, France}}

  \date{\today}
\begin{abstract}
We analyze the correspondence between two formalisms describing an interacting Bose gas, namely the standard Feynman diagrammatic expansion on the one hand, and the hierarchy equations for the imaginary-time Green functions on the other hand. We show that the Hartree-Fock approximation, as well as its first corrections at low density derived by Baym et al.  [Eur. Phys. J. B, \textbf{24}, pp. 107-124 (2001)], can be equivalently formulated in both formalisms. Within the hierarchy approach, the two-body correlations ${\cal G}^{(2,\rm T)}$ are expressed at lowest order in the interactions in terms of the full one-body Green function. This sheds light on the physical content of the corrections to the Hartree-Fock theory. Moreover, this representation of ${\cal G}^{(2,\rm T)}$ can be extended to higher orders,  opening  the way to systematic calculations of further corrections to the ideal critical temperature when the density increases. 
\end{abstract}
%\pacs{}
\maketitle

%\tableofcontents

%===============================================

%===============================================
\section{Introduction}
Bose Einstein (BE) condensation is an effect first predicted for an ideal Bose gas. Under a critical temperature or above a critical density, a condensate which correspond to a macroscopic occupancy of the lowest energy state, appears.  This phase transition is particular because it is due to the quantum statistics of the bosons and does not require interactions as it is usual in the context of phase transitions. We have thus to face the paradoxical question of the effects of interactions on this phase transition. In particular, we focus on a dilute gas with repulsive interactions which can hinder the condensation.

When considering an interacting Bose gas the first step is to study the strict mean-field (MF) case where interactions are taken with infinite range and vanishing amplitude \cite{PZ2004, BLS1984}. The MF hamiltonian was proved to predict that the BE transition persists. The MF transition displays the same structure as in the ideal case. The relation between the density and the chemical potential is modified by a density-dependent shift of the chemical potential of the interacting Bose gas. Hence, the density is determined, at given $T$ and $\mu$, by a self-consistent equation which gives rise to the same critical density $\rho_{c}$ as for the ideal gas. For $\rho > \rho_{c}$, a macroscopic fraction of the particles with density $(\rho - \rho_{c})$ is condensated in the lowest energy level.

The next step is to consider a more sophisticated mean-field approximation, the well-known Hartree-Fock (HF) approximation where one introduces an effective one-body potential. At fixed $T$ and $\mu$, the HF density is determined by solving two coupled self-consistent equations for the density and the effective potential. It has been shown that for sufficiently long-range but finite interaction, the critical point is erased and the Bose-Einstein condensation breaks down within the HF approximation~\cite{APS2019}. However, for short-range interactions, the critical density $\rho_{c}(T)$ still exists, and the nature of the corresponding transition is also unchanged. 

The HF approximation illustrates a possible consequence of the presence of interactions, namely the breakdown of the BE condensation. Thus a central question concerns the existence, nature and location of the transition beyond HF theory. Previous analytical works \cite{LY1958, GKW1960, H1999,T1982, BBH1999, BBH2001} have predicted, by using various approximations, different and sometimes incompatible interaction-induced shifts for the transition temperature. The upstream question of the nature of the transition, in particular the decay rate of the off-diagonal long-range order (ODLRO) in the one-body density matrix, is still not fully resolved. In view of this open question, we consider an interacting Bose gas in the normal phase, without condensate, and the issue is to determine whether a critical point with ODLRO does appear and its associated decay rate. Approaches where the presence of a condensate is assumed, like Hartree-Fock-Bogoliubov and Gross-Pitaevski theories, cannot answer this problem \cite{ZB2001,RS2013}. Baym et al. \cite{BBH2001} have examined the Bose-Einstein transition for a dilute Bose gas with short-range interactions: they take into account the a priori first correction to the HF approximation by keeping a suitable class of diagrams in the Feynman expansion. They have found that the transition persists and that the analytical properties of the critical point are modified with respect to the ideal and mean-field (MF and HF) cases.

The main purpose of the present paper is to shed light on the link between Feynman diagrammatic expansion and the hierarchy equations for the imaginary time Green functions. This correspondance is systematically established for the successive approximations that have been considered, including HF and its corrections derived in \cite{BBH2001}. Our analysis provides a better understanding of the physical content of the various approximation at hand. It opens also the way to systematic improvements of mean-field theories.

The paper is organized as follows. In Section 2, we introduce the model, and we briefly review the main tools involved in the Feynman diagrammatic expansion and in the hierarchy equations. In Section 3, we recall how the HF approximation is equivalently formulated in both formalisms. The mean-field nature of the HF approximation clearly emerges within the hierarchy approach since it amounts to factorize the two-body Green function as a product of one-body Green functions, {\it i.e.}\ to neglect two-body correlations induced by weak interactions. The hierarchy equations are well suited to study the BE condensation beyond HF, as developed in Section 4. It turns out that the full infinite hierarchy for the $n$-body Green functions contributes to the fully symmetrized lowest-order form of the two-body truncated Green function. Moreover, this form is shown to be equivalent to that of the proper self-energy introduced by Baym et al. We conclude in Section 5 with some comments and perspectives.

\section{Model and formalisms}

\subsection{Definitions}

We consider a system of $N$ ($N = 0, \cdots, \infty$) non-relativistic spinless bosons with mass $m$. We consider a two-body repulsive interaction $V(r)$ which we take positive, integrable, and spherically symmetric with

\begin{align}
\int d\bm{r} V(r) = 4 \pi \int_{0}^\infty dr r^2 V(r) = a.
\end{align} 
Note that in the MF approach, the energy shift of the kinetic energy is $a\rho$ where $\rho$ is the particle density \cite{BLS1984,PZ2004}.

The Hamiltonian of the system is
\begin{align}
\hat{\mathcal{H}} = - \sum \limits_{i=1}^N \frac{\hbar^2}{2m} \bm{\Delta}_{i} + \sum\limits_{i < j = 1}^N V(\abs{\bm{r}_{i} - \bm{r}_{j} }).
\end{align}

Here we consider a system in the Grand-canonical ensemble at fixed non-zero temperature $T$ and chemical potential $\mu$. We assume once for all that the thermodynamic limit has been taken, and that the resulting infinite system is in a fluid state, invariant under translations and rotations. All local quantities will refer to the infinite system, so they are implicitly assumed to be defined through the usual thermodynamic limit procedure defined by $\lim \limits_{\Lambda \to \infty}\biggr\rvert_{\mu, \beta = const}$ where $\Lambda$ is the volume of the occupied region of space and all positions are fixed infinitely far from the boundaries. We must then take the statistical Grand-canonical ensemble average (GC) in addition to the quantum expectation value. In the Grand-canonical ensemble, the relevant object is the partition function
\begin{align}
\mathcal{Z} = \Tr(\exp(-\beta (\hat{\mathcal{H}}-\mu\hat{\mathcal{N}})))
\end{align}
where the trace operator refers to a summation over all the possible states, $\beta = 1/k_{B} T$, and $\hat{\mathcal{N}}$ is the operator that counts the number of particles. The average value of an operator $\hat{\mathcal{A}}$ for non-zero temperature is
\begin{align}
\expval{A}_{GC} = \frac{ \Tr(\exp(-\beta (\hat{\mathcal{H}}-\mu\hat{\mathcal{N}}))\hat{\mathcal{A}})}{ \Tr(\exp(-\beta (\hat{\mathcal{H}}-\mu\hat{\mathcal{N}})))}  
= \frac{1}{\mathcal{Z} } \sum\limits_{N=0}^{\infty} \sum\limits_{n} \bra{\Psi_{n}^N}\exp(-\beta (\hat{\mathcal{H}}-\mu\hat{\mathcal{N}})) \hat{\mathcal{A}}\ket{\Psi_{n}^N}
\end{align}

The BE condensation is characterized by the appearance of an Off Diagonal Long Range Order (ODLRO) in the one-body density matrix~\cite{P1951,PO1956, Y1962}
\begin{align}
n^{(1)}(\bm{r},\bm{r'}) = \expval{\hat{\Psi^{\dagger}}(\bm{r}) \hat{\Psi}(\bm{r'})}_{\rm GC}
\la{eq:densitymatrix}
\end{align}
that we express in terms of the field operators $\hat{\Psi}(\bm{r})$ and $\hat{\Psi^{\dagger}}(\bm{r})$ which respectively annihilates and creates one particle at position $\bm{r}$ and satisfy canonical commutation relations. The limit of eq.~\eqref{eq:densitymatrix} when $| \bm{r} - \bm{r}'|  \to \infty$ can be seen as the order parameter of the transition
\begin{align}
\lim\limits_{| \bm{r} - \bm{r}'|  \to \infty} n^{(1)}(\bm{r},\bm{r'}) =
     \begin{cases}
      &0  \quad\textnormal{for\; $\rho<\rho_{c}$ (normal phase)}\\
      & \sim \displaystyle \frac{1}{| \bm{r} - \bm{r}'|^s} \quad \textnormal{with some $s$ at $\rho=\rho_{c}$}\\
      &\rho_{c} \quad \textnormal{for\; $\rho>\rho_{c}$ (condensate density $\rho-\rho_{c})$.}
    \end{cases}       
\end{align}
The appearance of ODLRO signals the onset of BE condensation. It provides an order parameter that does not assume independent quasiparticles. 

As the signature of the condensation is in the one-body density matrix, it is useful to introduce the one-body imaginary-time Green function
\begin{align}
\mathcal{G}^{(1)}(1 \mid 2) = \mathcal{G}^{(1)}(\bm{r_{1}},\tau_{1} \mid \bm{r_{2}}, \tau_{2}) = \expval{\hat{T}_{\tau} \hat{\Psi}_{\tau_{1}}(\bm{r_{1}}) \hat{\Psi}^{\dagger}_{\tau_{2}}(\bm{r_{2}})}_{GC}.
\end{align}
The operator $\hat{T}_{\tau}$ orders imaginary-time evolved operators,
\begin{align}
\hat{A}_{\tau} = \exp{\tau(\hat{\mathcal{H}}-\mu \hat{\mathcal{N}})} \hat{A} \exp{-\tau(\hat{\mathcal{H}}-\mu \hat{\mathcal{N}})},
\end{align}
with decreasing times from the left to the right, without changing sign of the corresponding product. The one-body Green function describes the propagation of one particle from $\bm{r_{1}}$ at imaginary-time $\tau_{1}$ to $\bm{r_{2}}$ at $\tau_{2}$. By construction, the Green function $\mathcal{G}^{(1)}(1 \mid 2)$ is $\beta\hbar$ periodic in each of these variables $\tau_{i}$ and commutation relations imply a discontinuity for equal imaginary-time $\tau_{1} = \tau_{2}$
\begin{align}
\mathcal{G}^{(1)}(1 \mid 2_{1^-}) - \mathcal{G}^{(1)}(1 \mid 2_{1^+}) = \delta(\bm{r_{1}} - \bm{r_{2}})
\end{align}
with $2_{1^-} = (\bm{r_{2}}, \tau_{1}^-)$ and $\tau^{\pm} =  \lim \limits_{\epsilon \to 0^+} (\tau \pm \epsilon)$.
For almost equal time $\tau_{2}=\tau_{1}^+$, the one-body Green function
\begin{align}
 \mathcal{G}^{(1)}(1 \mid 2_{1^+}) = n^{(1)}(\bm{r_{2}, \bm{r_{1}}}) = n^{(1)}(\abs{\bm{r_{2}- \bm{r_{1}}}})
\end{align}
provides the density matrix $n^{(1)}$.

\subsection{Feynman diagrams and self-energy}

A common procedure to determine the one-body Green function is based on the Feynman's perturbative diagrammatic expansion \cite{FW} in powers of the interaction, where the reference is the ideal Green function. Precise diagrammatic rules can be found in Ref.~\cite{FW}. Then, one can derive various approximations by summing particular classes of diagrams in the infinite series for $\mathcal{G}^{(1)}$,
\begin{equation}
{\setstretch{1.0}
\centering
\begin{tikzpicture}[baseline={([yshift=-.5ex]current bounding box.center)},vertex/.style={anchor=base}]

	\draw[->, line width = 2] (0,0)--(0,1);
	\draw[line width = 2] (0,1)--(0,2);
	\draw (0,0) node[below]{1};
	\draw (0,2) node[above]{2};
	\draw (0,-1) node[below,text width=2cm, align=center]{
Green function $\mathcal{G}^{(1)}$};
	
	\draw (1,1) node{$=$};
	
	\draw[->] (2,0)--(2,1);
	\draw[] (2,1)--(2,2);
	\draw (2,0) node[below]{1};
	\draw (2,2) node[above]{2};
	\draw (2,-1) node[below,text width=2.2cm, align=center]{Ideal Green function $\mathcal{G}^{(1)}_{0}$};
	
	\draw (3,1) node{$+$};
	
	\draw[->] (4,0)--(4,0.5);
	\draw[] (4,0.5)--(4,1);
	\draw (4,0) node[below]{1};
	\draw (4,2) node[above]{2};
	\draw[decorate,decoration={snake, segment length=5, amplitude=1}](4,1)--(5,1);
	\draw (5.5,1) circle (0.5);
	\draw[->] (6,1)--(6,0.99);
	\draw[->] (4,1)--(4,1.5);
	\draw[] (4,1.5)--(4,2);

	\draw (7,1) node{$+$};
	
	\draw[->] (8,0)--(8,0.33);
	\draw[] (8,0.33)--(8,0.66);
	\draw[->] (8,0.66)--(8,0.99);
	\draw[] (8,0.99)--(8,1.32);
	\draw[->] (8,1.32)--(8,1.66);
	\draw[] (8,1.66)--(8,2);
	\draw (8,0) node[below]{1};
	\draw (8,2) node[above]{2};
	\draw[decorate, decoration={snake, segment length=5, amplitude=1}] (8,0.66) to[bend left=-90] (8,1.32);
	
	\draw (9,1) node{$+$};
	
	\draw[->] (11,0)--(11,0.33);
	\draw[] (11,0.33)--(11,0.66);
	\draw[->] (11,0.66)--(11,0.99);
	\draw[] (11,0.99)--(11,1.32);
	\draw[->] (11,1.32)--(11,1.66);
	\draw[] (11,1.66)--(11,2);
	\draw[decorate, decoration={snake, segment length=5, amplitude=1}] (11,0.66) to[bend left=-90] (11,1.32);
	\draw[decorate,decoration={snake, segment length=5, amplitude=1}](10.5,1)--(11,1);
	\draw (10,1) circle (0.5);
	\draw[->] (9.5,1)--(9.5,1.01);
	\draw (11,0) node[below]{1};
	\draw (11,2) node[above]{2};
	
	\draw (12,1) node{$+$};
	
	\draw (13,1) node{$\cdots$};
	
	\draw[decorate,decoration={snake, segment length=5, amplitude=1}](7,-1.5)--(9,-1.5); 
	\draw (7,-1.5) node[left]{3};
	\draw (9,-1.5) node[right]{4};
	\draw (8,-2) node[below, align=center]{Interaction line between 3 and 4};
\end{tikzpicture}
}
%\caption{Pertubative expansion of $\mathcal{G}^{(1)}$ in terms of $\mathcal{G}^{(1)}_{0}$ and potential}
	\label{Fig:diag_series}
\end{equation}
The wavy line represents the interaction $V(\bm{i} \mid \bm{j}) = V(|\bm{r}_i - \bm{r}_j|) \delta(\tau_i - \tau_j)$. The first term in this series is the one-body Green function of the ideal gas, which is given by
\begin{align}
\mathcal{G}_{0}^{(1)}(1 \mid 2) &= \int \frac{d\bm{k}}{(2\pi)^3} \exp(-i\bm{k}\cdot(\bm{r_{1}}-\bm{r_{2}})) \cdot \frac{\exp(-(\tau_{1}-\tau_{2})[\epsilon(\bm{k})-\mu ])}{\exp(\beta [\epsilon(\bm{k})-\mu ])-1}\nonumber \\
&\cdot [  \theta(\tau_{2}-\tau_{1}) + \exp(\beta [\epsilon(\bm{k})-\mu ]) \theta(\tau_{1}-\tau_{2})  ]
\label{eq:freeprop}
\end{align}
where $\epsilon(\bm{k}) = \hbar^2 \bm{k}^2 / (2m)$ is the kinetic energy ($0 \leq \tau_{1}, \tau_{2} < \beta$) \cite{FW}.
One introduces a new quantity, the self energy $\Sigma$, by the graphical relation
\begin{equation}
\centering
\begin{tikzpicture}[baseline={([yshift=-.5ex]current bounding box.center)},vertex/.style={anchor=base}]

	\draw[->, line width = 2] (0,0)--(0,1);
	\draw[line width = 2] (0,1)--(0,2);
	
	\draw (1,1) node{$=$};
	
	\draw[->] (2,0)--(2,1);
	\draw[] (2,1)--(2,2);
	
	\draw (3,1) node{$+$};
	
	\draw[->] (4,0)--(4,0.25);
	\draw[] (4,0.25)--(4,0.5);
	\draw (4,1) circle (0.5);
	\draw (4,1) node{$\Sigma$};
	\draw[->] (4,1.5)--(4,1.75);
	\draw[] (4,1.75)--(4,2);
\end{tikzpicture}
%\caption{Self-energy definition}
\label{Fig:SE_def}
\end{equation}
which corresponds to the equation
\begin{align}
 \mathcal{G}^{(1)}(1 \mid 2) =   \mathcal{G}_{0}^{(1)}(1 \mid 2) + \int d\bm{3} d\bm{4} \quad \mathcal{G}_{0}^{(1)}(1 \mid 3) \cdot \Sigma(3 \mid 4) \cdot \mathcal{G}_{0}^{(1)}(4 \mid 2)
 \label{eq:partial_Dyson}
\end{align}
where integration over a point $\bm{i}=(\bm{r}_i, \,\tau_i)$ means $\int d\bm{i} = \int d\bm{r}_{i} \int_{0}^{\beta\hbar} \tau_{i}$. The self-energy $ \Sigma(3\mid4)$  accounts for the interactions, {\it i.e.} all diagrams in eq.~\eqref{Fig:diag_series} apart from the first one. As many diagrams in the self-energy can be built by using the same elementary blocks, it is useful to define the proper self-energy $ \Sigma^{\star}(3 \mid 4)$ as the sum of all self-energy diagrams that cannot be separated into two parts by cutting only one propagator line $\mathcal{G}_{0}^{(1)}$. Then, from the proper self-energy, one can generate all diagrams in $\Sigma$ by considering convolutions chains of arbitrary lengths made up of proper self-energies connected by propagators $\mathcal{G}_{0}^{(1)}$:
\begin{align}
\Sigma(1 \mid 2) = \Sigma^{\star}(1 \mid 2) + \int d\bm{3}  d\bm{4} \Sigma^{\star}(1 \mid 3) \mathcal{G}_{0}^{(1)} (3 \mid 4)  \Sigma^{\star}(4 \mid 2) + \cdots 
\label{II.15}
\end{align}
Inserting eq.~\eqref{II.15} into (\ref{eq:partial_Dyson}) makes the Green function emerge in the right hand side and leads to the Dyson equation \cite{B1958}
\begin{align}
\mathcal{G}^{(1)}(1 \mid 2) = \mathcal{G}_{0}^{(1)}(1 \mid 2) + \int d\bm{3} d\bm{4} \quad \mathcal{G}_{0}^{(1)}(1 \mid 3) \cdot \Sigma^{\star}(3 \mid 4) \cdot \mathcal{G}^{(1)}(4 \mid 2) \label{eq:Dyson}
\end{align}
The Hamiltonian is time independent and the system is supposed uniform. One assumes that the spatial Fourier transform exist and one takes the spatial Fourier transform of the Dyson equation (\ref{eq:Dyson}). As the Green functions are $\beta \hbar$ periodic in their imaginary-time arguments, they can be expressed as Fourier series over Matsubara frequencies $\omega_n = 2\pi n /(\beta \hbar)$, providing
\begin{align} \notag
\mathcal{G}^{(1)}(1 \mid 2) &= \mathcal{G}^{(1)}(\bm{r_{1}},\bm{r_{2}};\tau = \tau_{2}-\tau_{1})  = \frac{1}{\beta \hbar} \sum_{n} \exp(i \omega_{n}\tau) \tilde{\mathcal{G}}^{(1)}(\bm{r_{1}},\bm{r_{2}}; \omega_{n}) 
\\
&= \frac{1}{\beta \hbar} \int \frac{d \bm{k}}{(2\pi)^3} \exp(i \bm{k}\cdot (\bm{r_{1}}-\bm{r_{2}}))\sum_{n} \exp( i \omega_{n}\tau) \hat{\tilde{\mathcal{G}}}^{(1)}(\bm{k}; \omega_{n})
\end{align}
where
\begin{equation}
\tilde{\mathcal{G}}^{(1)}(\bm{r_{1}},\bm{r_{2}}; \omega_{n})  =  \int\limits_{0}^{\beta \hbar} \exp(-i \omega_{n} \tau)\mathcal{G}^{(1)}(\bm{r_{1}},\bm{r_{2}};\tau).
\end{equation}
In the frequency domain (both spatial and temporal), the Dyson equation becomes
\begin{align}
\hat{\tilde{\mathcal{G}}}^{(1)}(\bm{k}; \omega_{n})= \hat{\tilde{\mathcal{G}}}_{0}^{(1)}(\bm{k}; \omega_{n}) + \hat{\tilde{\mathcal{G}}}_{0}^{(1)}(\bm{k}; \omega_{n}) \cdot \hat{\tilde{\Sigma}}^{\star}(\bm{k}; \omega_{n}) \cdot \hat{\tilde{\mathcal{G}}}^{(1)}(\bm{k}; \omega_{n}). \label{eq:T_Dyson}
\end{align}
From this algebraic equation, one can express the one-body Fourier and Matsubara transformed Green function in terms of the transformed ideal Green function and proper self-energy :
\begin{align}
\hat{\tilde{\mathcal{G}}}^{(1)}(\bm{k}; \omega_{n}) = \frac{1}{\hat{\tilde{\mathcal{G}}}_{0}^{(1)}(\bm{k}; \omega_{n})^{-1} - \frac{1}{\hbar} \hat{\tilde{\Sigma}}^{\star}(\bm{k}; \omega_{n}) } = \frac{1}{ -i \omega_{n} + \frac{1}{\hbar} \[ \epsilon(\bm{k}) - \mu -  \hat{\tilde{\Sigma}}^{\star}(\bm{k}; \omega_{n}) \]}
\la{eq:dysonfulltransformed}
\end{align}
where we have used, in the second equality,
\begin{align}
\hat{\tilde{\mathcal{G}}}^{(1)}_{0}(\bm{k}; \omega_{n}) = \frac{1}{-i \omega_{n} +\frac{1}{\hbar} \[ \epsilon(\bm{k}) - \mu \]}.
\label{eq:Gid1(k,omega)}
\end{align}

\subsection{Hierarchy equation for the $n$-body Green functions}

Since the Green functions are expressed in terms of field operators, one can use the equations of motion for such operators and the second-quantization representation of $\hat{\mathcal{H}}$ and $\hat{\mathcal{N}}$,
\begin{align}
\hat{\mathcal{H}} = - \frac{\hbar^2}{2m} \int d\bm{r} \hat{\Psi}^{\dagger}(\bm{r}) \Delta  \hat{\Psi}(\bm{r}) + \frac{1}{2} \int d\bm{r} \int d\bm{r'} \hat{\Psi}^{\dagger}(\bm{r})\hat{\Psi}^{\dagger}(\bm{r'}) V(\bm{r} - \bm{r'}) \hat{\Psi} (\bm{r}) \hat{\Psi}(\bm{r'}) 
\end{align}
\begin{align}
\hat{\mathcal{N}} = \int d\bm{r} \hat{\Psi}^{\dagger}(\bm{r})  \hat{\Psi}(\bm{r}),
\end{align}
to obtain
\begin{align}
\hbar \frac{\partial}{\partial \tau_{1}} \hat{\Psi}_{\tau_{1}}(\bm{r_{1}}) &= \comm{\hat{\mathcal{H}}-\mu\hat{\mathcal{N}}}{\hat{\Psi}(\bm{r_{1}})}_{\tau_{1}}\nonumber \\
&= \frac{\hbar^2}{2m} \Delta_{1} \hat{\Psi}_{\tau_{1}}(\bm{r_{1}}) + \mu \hat{\Psi}_{\tau_{1}}(\bm{r_{1}}) - \int d\bm{r_{3}} \hat{\Psi}^{\dagger}_{\tau_{1^+}}(\bm{r_{3}}) V(\abs{\bm{r_{3}}-\bm{r_{1}}}) \hat{\Psi}_{\tau_{1}}(\bm{r_{3}})\hat{\Psi}_{\tau_{1}}(\bm{r_{1}}) \label{eq:motionfieldoperator}
\end{align}
 From (\ref{eq:motionfieldoperator}), one establishes an equation of motion for the imaginary-time one-body Green function called the first hierarchy equation
 \begin{align}
 \nonumber
 \hbar \frac{\partial}{\partial\tau_{1}} \mathcal{G}^{(1)}(1 \mid 2) =\; &  (\frac{\hbar^2}{2m} \Delta_{1}+\mu) \mathcal{G}^{(1)}(1 \mid 2) + \delta(\bm{r_{1}}-\bm{r_{2}}) \delta(\tau_{1}-\tau_{2}) \\
 & - \int d\bm{r_{3}}  V(\abs{\bm{r_{3}}-\bm{r_{1}}}) \mathcal{G}^{(2)}(1, 3_{1} \mid 3_{1^+},2). 
 \label{eq:firsthierarchyeq}
 \end{align}
On the right hand side, there are different terms: a standard kinetic term, a $\delta$-term which takes into account the discontinuity at $\tau_{1} = \tau_{2}$ and finally an interaction term which involves the two-body Green function due to pair-wise interaction where the $\tau_{1^+}$ imaginary time is due to the $T$-product. One can also introduce the $n$-body Green function
 \begin{align}
  \mathcal{G}^{(n)}(1, 3, \cdots, 2n-1 \mid 2, 4, \cdots, 2n) = \expval{\hat{T}_{\tau} \hat{\Psi}_{\tau_{1}}(\bm{r_{1}}) \cdots \hat{\Psi}_{\tau_{2n-1}}(\bm{r_{2n-1}})  \hat{\Psi}^{\dagger}_{\tau_{2}}(\bm{r_{2}}) \cdots \hat{\Psi}^{\dagger}_{\tau_{2n}}(\bm{r_{2n}})}_{GC}
 \end{align}
 By the same method, we can derive an equation of motion for the two-body Green function involving the three-body Green function
\begin{align}
\hbar \frac{\partial}{\partial\tau_{1}} \mathcal{G}^{(2)}(1,3 \mid 2,4) &= ( \frac{\hbar^2}{2m} \Delta_{1}+\mu) \mathcal{G}^{(2)}(1 ,3\mid 2,4)  - \int d\bm{r_{5}}  V(\abs{\bm{r_{5}}-\bm{r_{1}}}) \mathcal{G}^{(3)}(1, 3, 5_{1} \mid 5_{1^+},2,4) \nonumber \\
&+ \delta(\bm{r_{1}}-\bm{r_{2}}) \delta(\tau_{1}-\tau_{2}) \mathcal{G}^{(1)}(3 \mid 4) + \delta(\bm{r_{1}}-\bm{r_{4}}) \delta(\tau_{1}-\tau_{4}) \mathcal{G}^{(1)}(3 \mid 2)  \label{eq:second_eq}
\end{align}
In the following, we show how various approximations, including HF and corrections to HF, can be equivalently formulated by using both formalisms.

\section{Hartree-Fock}
As explained before, one can determine the Green function $\mathcal{G}^{(1)}$ from any given proper self-energy. Using a simple proper self-energy, the Dyson equation amounts to take into account a large number of diagrams in the Feynman expansion. Nevertheless, there is an infinite number of proper self-energy diagrams, up to arbitrary high orders in the interaction.  The proper self-energy diagrams at first order (one interaction line) are
\begin{equation}
\begin{tikzpicture}[scale=0.75,baseline={([yshift=-.5ex]current bounding box.center)},vertex/.style={anchor=base}]
\draw (2,1) node{$\Sigma^{\star}_{(1)}$};
	
	\draw (3,1) node{$=$};

	\draw[decorate,decoration={snake, segment length=5, amplitude=1}](4,1)--(4.75,1);
	\draw (5.5,1) circle (0.75);
	\draw[->] (6.25,1)--(6.25,0.99);
	
	\draw (7,1) node{$+$};

	\draw[->] (8,0)--(8,1);
	\draw (8,1)--(8,2);
	\draw[decorate, decoration={snake, segment length=5, amplitude=1}] (8,0) to[bend left=-90] (8,2);
\end{tikzpicture}
%\caption{All first order in potential proper self-energy diagrams (one interaction line)}
\label{Fig:first_order_proper}
\end{equation}
and those at second order are
\begin{equation}
\begin{tikzpicture}[scale=0.75,baseline={([yshift=-.5ex]current bounding box.center)},vertex/.style={anchor=base}]

\draw (-3,-8) node{$\Sigma^\star_{(2)}$};

\draw(-2,-8) node{$=$};

\draw[decorate,decoration={snake, segment length=5, amplitude=1}](-1,-8)--(0,-8);
\draw (0.5,-8) circle (0.5);
\draw[->](1,2.001-10)--(1,1.999-10);
\draw[decorate,decoration={snake, segment length=5, amplitude=1}](0.5,-7.5)--(0.5,-8.5);

\draw(2,-8) node{$+$};

\draw[decorate,decoration={snake, segment length=5, amplitude=1}](3,-8)--(4,-8);
\draw (4.5,-8) circle (0.5);
\draw[->](4.499,-7.5)--(4.5001,-7.5);
\draw[<-](4.499,-8.5)--(4.5001,-8.5);
\draw[decorate,decoration={snake, segment length=5, amplitude=1}](5,-8)--(6,-8);
\draw (6.5,-8) circle (0.5);
\draw[->](7,2.001-10)--(7,1.999-10);

\draw(8,-8) node{$+$};

\draw[](9,-8)--(9,-6.5);
\draw[->](9,-9.5)--(9,-8);
\draw[decorate,decoration={snake, segment length=5, amplitude=1}] (9,-9.5) to[bend left=-90] (9,-6.5);
\draw[decorate,decoration={snake, segment length=5, amplitude=1}] (9,-9) to[bend left=-90] (9,-7);

\draw(11,-8) node{$+$};

\draw[](13,-7.25)--(13,-6.5);
\draw[->](13,-8.75)--(13,-7.25);
\draw[->](13,-9.5)--(13,-8.75);
\draw[decorate,decoration={snake, segment length=5, amplitude=1}] (13,-9.5) to[bend left=90] (13,-6.5);
\draw[decorate,decoration={snake, segment length=5, amplitude=1}](13,-8)--(14,-8);
\draw (14.5,-8) circle (0.5);
\draw[->](15,2.001-10)--(15,1.999-10);

\draw(-2,-12) node{$+$};

\draw[](-1,-12)--(-1,-11);
\draw[->](-1,-13)--(-1,-12);
\draw[decorate,decoration={snake, segment length=5, amplitude=1}] (-1,-11)--(0,-11);
\draw(0,-13) to[bend left=-45] (0,-11);
\draw(0,-13) to[bend left=45] (0,-11);
\draw[decorate,decoration={snake, segment length=5, amplitude=1}] (-1,-13)--(0,-13);
\draw[->] (-0.41,-12.01)--(-0.41,-11.99);
\draw[<-] (0.41,-12.01)--(0.41,-11.99);

\draw(2,-12) node{$+$};

\draw[decorate,decoration={snake, segment length=5, amplitude=1}] (4,-10.5) to[bend left = 90](4,-13);\draw[decorate,decoration={snake, segment length=5, amplitude=1}] (4,-11) to[bend left = -90](4,-13.5);
\draw[->] (4,-13.5)--(4,-12);
\draw (4,-12)--(4,-10.5);
\end{tikzpicture}
%\caption{All second order in potential proper self-energy diagrams (two interaction lines)}
\label{Fig:second_order_proper}
\end{equation}
Even if the diagrams in $\Sigma^\star_{(2)}$ cannot be separated in two parts by cutting one propagator line, the first four are made with diagrams of $\Sigma^\star_{(1)}$ where free propagator lines are attached to another diagrams of $\Sigma^\star_{(1)}$. Hence, one can build proper self-energy diagrams at any order from these two first-order diagrams. In the counting of all these diagrams in the proper self-energy, the Green function $\mathcal{G}^{(1)}$ emerges and gives the graphical definition of the HF proper self-energy, namely
\begin{equation}
\begin{tikzpicture}[scale = 0.75,baseline={([yshift=-.5ex]current bounding box.center)},vertex/.style={anchor=base}]
	\draw (2.5,1) node[left]{$\Sigma^{\star}_{\rm{HF}}(\bm{3}, \bm{4})$};
	
	\draw (3,1) node{$=$};

	\draw[decorate,decoration={snake, segment length=5, amplitude=1}](4,1)--(4.75,1);
	\draw (4,1) node[below left]{$\bm{3}$};
	\draw (4,1) node[above left]{$\bm{4}$};
	\draw[ line width = 2] (5.5,1) circle (0.75);
	\draw (4.8,1) node[right]{$\bm{5}$};
	\draw[->, line width = 2] (6.25,1)--(6.25,0.99);
	
	\draw (7,1) node{$+$};

	\draw[->, line width = 2] (8,0)--(8,1);
	\draw[ line width = 2] (8,1)--(8,2);
	\draw (8,0) node[left]{$\bm{3}$};
	\draw (8,2) node[left]{$\bm{4}$};
	\draw[decorate, decoration={snake, segment length=5, amplitude=1}] (8,0) to[bend left=-90] (8,2);
	
	\draw [decorate, decoration = {brace,mirror}] (4,-0.2) --  (6.5,-0.2);
	\draw (5.25,-0.2) node[below]{$\Sigma^{\star}_{\rm{Hartree}}$};
	
	\draw [decorate, decoration = {brace,mirror}] (7.75,-0.20) --  (8.75,-0.2);
	\draw (8.25,-0.2) node[below]{$\Sigma^{\star}_{\rm{Fock}}$};
\end{tikzpicture}
%\caption{HF proper self-energy}
\label{Fig:HF_proper}
\end{equation}

The Hartree term contains a propagator loop corresponding to $\mathcal{G}^{(1)}(5,5^+) = n^{(1)}(\bm{r_{5}},\bm{r_{5}}) = \rho $ in the real space where $\rho$ is the particle number density. Then the integration on the point $\bm{r_{5}}$ gives the constant $a = \int d \bm{r_{5}} V(\abs{\bm{r_{5}}-\bm{r_{3}}})$. The Hartree proper self-energy contributes as $a\rho$, the strict mean-field energy, in the particle spectrum.
\begin{align}
\Sigma^{\star}_{\rm{Hartree}}(\bm{3},\bm{4}) = \delta(\bm{r_{4}}-\bm{r_{3}}) a \rho
\end{align}

In agreement with the rigorous proof \cite{BLS1984, PZ2004}, in the strict mean-field, only this Hartree term is expected to remain in the limit $\gamma \to 0$ of an infinite long-range Kac potential $V_{\gamma}(r) = \gamma^3 v(\gamma r)$ with $v(x)$ fixed once for all and $\int d\bm{x} v(x) = a$. It can be checked that all the Feynman diagrams beyond the Hartree proper self-energy indeed vanish when $\gamma \to 0$.

 Concerning the Fock term, it is the product of a propagator and of an interaction line. Since the interaction is at equal imaginary time and because the system is invariant under spatial translations and rotations, we assume the Green function has the same properties and the Fock proper self-energy is written 
\begin{align}
\Sigma^{\star}_{\rm{Fock}}(\bm{3},\bm{4}) =  V(\abs{\bm{r_{4}}-\bm{r_{3}}}) \cdot \mathcal{G}^{(1)}(3,4_{3^+}) = V(\abs{\bm{r_{4}}-\bm{r_{3}}}) \cdot n^{(1)}(\abs{\bm{r_{4}}-\bm{r_{3}}}) \equiv \phi(\abs{\bm{r_{4}}-\bm{r_{3}}})
\end{align}
The Dyson equation (\ref{eq:Dyson}) gives then the Hartree-Fock Green function
\begin{align}
\hat{\tilde{\mathcal{G}}}^{(1)}_{\rm{HF}}(\bm{k}; \omega_{n}) = \frac{1}{-i \omega_{n} + \frac{1}{\hbar} \[\epsilon(\bm{k}) - \mu +a\rho +\hat{\phi}(\abs{\bm{k}})\] }
\end{align}
whose form is similar to that of the ideal Green function. Note that the one-particle energy spectrum is shifted by the strict mean field energy, $a\rho$, and by an effective potential $\phi$ called the HF potential.
By summing over all Matsubara frequencies and by taking the inverse Fourier transform, one expresses the HF Green function as in the ideal case as 
\begin{align}
\mathcal{G}_{\rm{HF}}^{(1)}(1 \mid 2) &= \int \frac{d\bm{k}}{(2\pi)^3} \exp(-i\bm{k}\cdot(\bm{r_{1}}-\bm{r_{2}})) \cdot \frac{\exp(-(\tau_{1}-\tau_{2})\frac{1}{\hbar}[\epsilon(\bm{k})-\mu  +a\rho +\hat{\phi}(\abs{\bm{k}})])}{\exp(\beta [\epsilon(\bm{k})-\mu  +a\rho +\hat{\phi}(\abs{\bm{k}}) ])-1} \nonumber \\
&\cdot [  \theta(\tau_{2}-\tau_{1}) + \exp(\beta [\epsilon(\bm{k})-\mu  +a\rho +\hat{\phi}(\abs{\bm{k}}) ]) \theta(\tau_{1}-\tau_{2})  ] \label{eq:HF_Green_function}
\end{align}
The particle number density $\rho = \mathcal{G}^{(1)}(\bm{1} \mid \bm{1^+})$ is the Green function evaluated at a special configuration of position and imaginary times, so it is determined self-consistently. In the same way, the HF potential is defined from the Green function which depends itself on the HF potential. In the Hartree-Fock approximation, one has to solve these two self-consistent integral equations
\begin{align}
\rho&= \int \frac{d\bm{k}}{(2\pi)^3}   \frac{1}{\exp(\beta [\epsilon(\bm{k})-\mu  +a\rho +\hat{\phi}(\abs{\bm{k}}) ])-1} \label{eq:self_density}\\
\phi(\abs{\bm{q}})&= \int \frac{d\bm{k}}{(2\pi)^3}   \frac{\hat{V}(\abs{\bm{q}-\bm{k}})}{\exp(\beta [\epsilon(\bm{k})-\mu  +a\rho +\hat{\phi}(\abs{\bm{k}}) ])-1} \label{eq:self_potential}
\end{align}
where $\rho$ corresponds to evaluating eq.~\eqref{eq:HF_Green_function} at equal positions and for almost equal times $\tau_{2} = \tau_{1}^+$ and $\hat{\phi}$ to a convolution of $\hat{\mathcal{G}}^{(1)}_{\rm HF}$ and $\hat{V}$. 
Diagrammatically, the HF approximation corresponds to a ``first order self-consistent approximation'' in which one keeps a special class of diagram. Neglecting all the other diagrams is rather questionnable. 

Interestingly, one can easily recover the HF approximation in the hierarchy formalism. For free particles, the first hierarchy equation reduces to
 \begin{align}
 \hbar \frac{\partial}{\partial\tau_{1}} \mathcal{G}^{(1)}_{0}(1 \mid 2) =  (\frac{\hbar^2}{2m} \Delta_{1}+\mu) \mathcal{G}^{(1)}_{0}(1 \mid 2) + \delta(\bm{r_{1}}-\bm{r_{2}}) \delta(\tau_{1}-\tau_{2}) \label{eq:freepropeq}
 \end{align}
 The spatial invariance under translation suggests to Fourier transform this equation and solve it by the variation of parameters method. Imaginary time periodicity is then imposed by boundary conditions and leads to the ideal Green function~\eqref{eq:freeprop} \cite{AP2011}.
 The complete first hierarchy equation involves the two-body Green function, $\mathcal{G}^{(2)}(\bm{1}, \bm{3} \mid \bm{2}, \bm{4})$, corresponding to the propagation of two particles from input points $(\bm{1}, \bm{3})$ to output points $(\bm{2}, \bm{4})$. An elementary way to propagate two particles is to propagate them separately, {\it i.e.}to propagate each particle from one of the input points to one of the output points without any interaction between them. This amounts to replace $\mathcal{G}^{(2)}$ by $\mathcal{G}^{(1)} \mathcal{G}^{(1)}$. This leads to the introduction of the two-body truncated Green function $\mathcal{G}^{(2,\rm T)}$ which reflects correlations induced by the interactions, via the decomposition
 \begin{align}
 \mathcal{G}^{(2)}(\bm{1}, \bm{3} \mid \bm{2}, \bm{4}) = \mathcal{G}^{(1)}(1 \mid 2) \cdot \mathcal{G}^{(1)}(3 \mid 4) + \mathcal{G}^{(1)}(1 \mid 4) \cdot \mathcal{G}^{(1)}(3 \mid 2) + \mathcal{G}^{(2,\rm T)}(\bm{1}, \bm{3} \mid \bm{2}, \bm{4}) \la{eq:decompo}
 \end{align}
 Setting $\mathcal{G}^{(2,\rm T)}=0$ amounts to neglect correlations induced by the interactions. Inserting the corresponding form of $\mathcal{G}^{(2)}$ into the first hierarchy equation \eqref{eq:firsthierarchyeq} leads to the now closed first hierarchy equation
 \begin{align}
  \nonumber
\hbar \frac{\partial}{\partial\tau_{1}} \mathcal{G}^{(1)}(1 \mid 2) &= (\frac{\hbar^2}{2m} \Delta_{1}+\mu-a\rho) \mathcal{G}^{(1)}(1 \mid 2) + \delta(\bm{r_{1}}-\bm{r_{2}}) \delta(\tau_{1}-\tau_{2}) \\
&- \int d\bm{r_{3}}  \underbrace{V(\abs{\bm{r_{3}}-\bm{r_{1}}}) n^{(1)}(\abs{\bm{r_{3}}-\bm{r_{1}}})}_{\phi(\abs{\bm{r_{3}}-\bm{r_{1}}})}\mathcal{G}^{(1)}(3_{1} \mid 2) \label{eq:HFhierar}
\end{align}
where one recognizes the HF potential $\phi$.

Taking the Fourier transform of this closed equation gives then the same equation than for an ideal gas with an energy spectrum shifted by the quantity $a\rho + \hat{\phi}(\bm{k})$. Thus one recovers the HF Green function~\eqref{eq:HF_Green_function} where the density and the HF potential are defined self-consistently by the equations (\ref{eq:self_density}) and (\ref{eq:self_potential}).

In the hierarchy formalism, the HF approximation clearly emerges as a mean-field approximation because it amounts to neglect two-particle correlations. In the following section, we study corrections to the HF approximation which incorporate two-particle correlations within the hierarchy formalism. Then we compare the resulting Green function with that derived beyond the HF expression within the Feynman diagrams \cite{BBH2001}.

\section{Beyond Hartree-Fock}

\subsection{Closure of the hierarchy at a given level}	\label{SctIVA}

To go beyond the HF approximation, we need to consider correlations induced by the interactions and so to derive a non-vanishing truncated two-body Green function $\mathcal{G}^{(2,\rm T)}$. Similarly to the decomposition (\ref{eq:decompo}) of the two-body Green function, the three-body Green function is decomposed into a truncated part and products of lower-order Green functions, namely
\begin{align}
\mathcal{G}^{(3)} =  \mathcal{G}^{(1)} \cdot \mathcal{G}^{(1)} \cdot \mathcal{G}^{(1)} + \mathcal{G}^{(1)} \cdot \mathcal{G}^{(2,\rm T)} + \mathcal{G}^{(3,\rm T)}.
\end{align}
Inserting this decomposition into the second hierarchy equation (\ref{eq:second_eq}), one obtains
\begin{alignat}{2}
\hbar \frac{\partial}{\partial\tau_{1}} \mathcal{G}^{(2,\rm T)} &= ( \frac{\hbar^2}{2m} \Delta_{1}+\mu) \mathcal{G}^{(2,\rm T)}  \nonumber \\
&- \mathcal{G}^{(1)} \int V\cdot\mathcal{G}^{(1)} \cdot \mathcal{G}^{(1)}
  \quad&\leftarrow\textnormal{2 terms of this type}\\
&- \int V \cdot \mathcal{G}^{(1)} \cdot \mathcal{G}^{(2,\rm T)} 
  &\leftarrow\textnormal{7 terms of this type} \nonumber \\
&- \int V \cdot \mathcal{G}^{(3,\rm T)}  
 &\leftarrow\textnormal{1 term of this type\phantom{s}} \nonumber
\end{alignat}
It is tempting to set $\mathcal{G}^{(3,\rm T)}=0$ in this equation, {\it i.e.} closing the hierarchy at the level $\mathcal{G}^{(2,\rm T)}$. Moreover, in addition to the two terms $\mathcal{G}^{(1)} \int V\cdot\mathcal{G}^{(1)} \cdot \mathcal{G}^{(1)}$, we can keep only the two terms of the form $\int V \cdot \mathcal{G}^{(1)} \cdot \mathcal{G}^{(2,\rm T)}$ that give $-a \rho \mathcal{G}^{(2,\rm T)}(\bm{1},\bm{3} \mid \bm{2}, \bm{4})$ and $- \int \phi(\mid \bm{r_{5}}- \bm{r_{1}} \mid)  \cdot \mathcal{G}^{(2,\rm T)} (\bm{3},\bm{5_{1}} \mid \bm{2}, \bm{4})$ respectively. This procedure is a closure of the second hierarchy equation. The corresponding truncated two-body Green function verifies an equation similar to that of the HF one-body Green function, but with two additional source terms involving the $\mathcal{G}^{(1)}$'s, {\it i.e.}
\begin{align}
\hbar \frac{\partial}{\partial\tau_{1}} \mathcal{G}^{(2,\rm T)}(\bm{1},\bm{3} \mid \bm{2}, \bm{4}) &= ( \frac{\hbar^2}{2m} \Delta_{1}+\mu -a \rho) \mathcal{G}^{(2,\rm T)}(\bm{1},\bm{3} \mid \bm{2}, \bm{4})\nonumber \\
&- \int \phi(\mid \bm{r_{5}}- \bm{r_{1}} \mid)  \cdot \mathcal{G}^{(2,\rm T)} (\bm{3},\bm{5_{1}} \mid \bm{2}, \bm{4}) \la{eq:G2T2sources} \\
&- \mathcal{G}^{(1)}(\bm{1} \mid \bm{2}) \int d \bm{r_{5}} V(\bm{r_{5}}-\bm{r_{1}})\cdot\mathcal{G}^{(1)}(\bm{3} \mid \bm{5_{1^+}}) \cdot \mathcal{G}^{(1)} (\bm{5_{1}} \mid \bm{4}) \nonumber\\
&- \mathcal{G}^{(1)}(\bm{1} \mid \bm{4}) \int d \bm{r_{5}} V(\bm{r_{5}}-\bm{r_{1}})\cdot\mathcal{G}^{(1)}(\bm{3} \mid \bm{5_{1^+}}) \cdot \mathcal{G}^{(1)} (\bm{5_{1}} \mid \bm{2})\nonumber \nonumber
\end{align}
By taking the spatial Fourier transforms, we can rewrite \eqref{eq:G2T2sources} as  
\begin{align}
&\hbar \frac{\partial}{\partial\tau_{1}} \hat{\mathcal{G}}^{(2,\rm T)}\left(\begin{matrix}
    \bm{k_{1}}\\[-0.5em]
    \tau_{1}
    \end{matrix}, 
    \begin{matrix}
    \bm{k_{3}}\\[-0.5em]
    \tau_{3}
    \end{matrix}
\Bigm\vert
\begin{matrix}
    \bm{k_{2}}\\[-0.5em]
    \tau_{2}
    \end{matrix}
,\begin{matrix}
   -\bm{k_{1}}-\bm{k_{2}}-\bm{k_{3}}\\[-0.5em]
    \tau_{4}
    \end{matrix} \right) = \\
    & [ -\frac{\hbar^2 k^2_{1}}{2m} +\mu -a \rho - \phi(\bm{k_{1}}) ] \hat{\mathcal{G}}^{(2,\rm T)}\left(\begin{matrix}
    \bm{k_{1}}\\[-0.5em]
    \tau_{1}
    \end{matrix}, 
    \begin{matrix}
    \bm{k_{3}}\\[-0.5em]
    \tau_{3}
    \end{matrix}
\Bigm\vert
\begin{matrix}
    \bm{k_{2}}\\[-0.5em]
    \tau_{2}
    \end{matrix}
,\begin{matrix}
   -\bm{k_{1}}-\bm{k_{2}}-\bm{k_{3}}\\[-0.5em]
    \tau_{4}
    \end{matrix} \right) \nonumber \\
&- [  \hat{V}(\bm{k_{1}}+\bm{k_{2}}) + \hat{V}(\bm{k_{2}}+\bm{k_{3}})  ] \hat{\mathcal{G}}^{(1)}(\bm{k_{2}} ; \tau_{1},\tau_{2}) \cdot \hat{\mathcal{G}}^{(1)}(\bm{k_{3}} ; \tau_{3},\tau_{1^+}) \cdot \hat{\mathcal{G}}^{(1)} (\bm{k_{1}}+\bm{k_{2}}+\bm{k_{3}} ; \tau_{1},\tau_{4})
\end{align}
This equation, as well as first equation of the hierarchy which involves this $\hat{\mathcal{G}}^{(2,\rm T)}$, are solved by the variation of parameters method, which first leads to an expression of $\mathcal{G}^{(2,\rm T)}$ in terms of $\mathcal{G}^{(1)}$, and then in a second step to a self-consistent equation for $\mathcal{G}^{(1)}$. In the first step, we find
\begin{align}
&\hat{\mathcal{G}}^{(2,\rm T)}\(\begin{matrix}
    \bm{k_{1}}\\[-0.5em]
    \tau_{1}
    \end{matrix}, 
    \begin{matrix}
    \bm{k_{3}}\\[-0.5em]
    \tau_{3}
    \end{matrix}
\Bigm\vert
\begin{matrix}
    \bm{k_{2}}\\[-0.5em]
    \tau_{2}
    \end{matrix}
,\begin{matrix}
   \bm{k_{4}}\\[-0.5em]
    \tau_{4}
    \end{matrix} \) =  - [  \hat{V}(\bm{k_{1}}+\bm{k_{2}}) + \hat{V}(\bm{k_{2}}+\bm{k_{3}}) ]  \nonumber \\
& \cdot \int\limits^{\beta\hbar}_{0} \frac{ds}{\hbar}\quad  \hat{\mathcal{G}}^{(1)}_{\rm HF}(\bm{k_{1}};0,s) \hat{\mathcal{G}}^{(1)}(\bm{k_{2}} ;s+\tau_{1},\tau_{2}) \cdot \hat{\mathcal{G}}^{(1)}(\bm{k_{3}} ; \tau_{3},s+\tau_{1}) \cdot \hat{\mathcal{G}}^{(1)} (\bm{k_{4}} ;s+\tau_{1},\tau_{4}).\label{eq:G2Tapprox}
\end{align}
where $\bm{k_{4} = - (\bm{k_{1}}+\bm{k_{2}}+\bm{k_{3}}})$.
It is useful to represent such equations with Feynman diagrams.
Graphically, the definition \eqref{eq:decompo} of $\mathcal{G}^{(2,\rm T)}$ is
\begin{equation}
\begin{tikzpicture}[scale=0.5, baseline={([yshift=-.5ex]current bounding box.center)},vertex/.style={anchor=base}]
	\draw[] (0,-2) to[bend left=-20] (0,2);
	\draw[] (2,-2) to[bend left=20] (2,2);
	\draw[] (0,-2)  to[bend left=40] (2,-2);
	\draw[] (0,2)  to[bend left=-40] (2,2);
	\draw (0, 2) node[right]{$\bullet$};
	\draw (2, 2) node[left]{$\bullet$};
	\draw (0, 2) node[left]{2};
	\draw (2, 2) node[right]{4};
	\draw (0, -2) node[left]{1};
	\draw (2, -2) node[right]{3};
	\draw [decorate, decoration = {brace,mirror}] (-1,-2.5) --  (3,-2.5);
	
	\draw (1,-3) node[below]{$\mathcal{G}^{(2)}(1,3 \mid 2,4)$};
	
	\draw (4, 0) node{$=$};
	
	\draw[] (6,-2) to[bend left=-20] (6,2);
	\draw[] (8,-2) to[bend left=20] (8,2);
	\draw[] (6,-2)  to[bend left=40] (8,-2);
	\draw[] (6,2)  to[bend left=-40] (8,2);
	\draw (6, 2) node[right]{$\bullet$};
	\draw (8, 2) node[left]{$\bullet$};
	\draw (6, 2) node[left]{2};
	\draw (8, 2) node[right]{4};
	\draw (6, -2) node[left]{1};
	\draw (8, -2) node[right]{3};
	\draw [decorate, decoration = {brace,mirror}] (5,-2.5) --  (9,-2.5);
	\draw (7,0) node{T};
	
	\draw (7,-3) node[below]{$\mathcal{G}^{(2,\rm T)}(1,3 \mid 2,4)$};
	
	\draw (9, 0) node{$+$};
	
	\draw[->, line width = 2] (10,-2)--(10,0);
	\draw[line width = 2] (10,-2)--(10,2);
	\draw (10, 2) node[above]{2};
	\draw (10, -2) node[below]{1};
	\draw[->, line width = 2] (11,-2)--(11,0);
	\draw[line width = 2] (11,-2)--(11,2);
	\draw (11, 2) node[above]{4};
	\draw (11, -2) node[below]{3};
	
	\draw (12, 0) node{$+$};
	
	\draw[->, line width = 2] (13,-2)--(13,0);
	\draw[line width = 2] (13,-2)--(13,2);
	\draw (13, 2) node[above]{4};
	\draw (13, -2) node[below]{1};
	\draw[->, line width = 2] (14,-2)--(14,0);
	\draw[line width = 2] (14,-2)--(14,2);
	\draw (14, 2) node[above]{2};
	\draw (14, -2) node[below]{3};
\end{tikzpicture}
\end{equation}
where one represents $\mathcal{G}^{(2,\rm T)}$ by a stretched sheet with the output points marked by a dot to distinguish them from input points.
Eq.~(\ref{eq:G2Tapprox}) can then be expressed diagrammatically by means of propagator (bold line: $\hat{\mathcal{G}}^{(1)}$, dash-dotted line: HF)
and interaction lines 
\begin{equation}
\centering
\begin{tikzpicture}[scale=0.5,baseline={([yshift=-.5ex]current bounding box.center)},vertex/.style={anchor=base}]
	\draw[] (6,-2) to[bend left=-20] (6,2);
	\draw[] (8,-2) to[bend left=20] (8,2);
	\draw[] (6,-2)  to[bend left=40] (8,-2);
	\draw[] (6,2)  to[bend left=-40] (8,2);
	\draw (6, 2) node[right]{$\bullet$};
	\draw (8, 2) node[left]{$\bullet$};
	\draw (6, 2) node[left]{2};
	\draw (8, 2) node[right]{4};
	\draw (6, -2) node[left]{1};
	\draw (8, -2) node[right]{3};
	\draw (7,0) node{T};

	\draw (9, 0) node{$=$};
	
	\draw[->,dashdotted, line width = 1.5] (10,-2)--(10,-1);
	\draw[dashdotted, line width = 1.5] (10,-2)--(10,0);
	\draw[->, line width = 2] (10,0)--(10,1);
	\draw[ line width = 2] (10,0)--(10,2);
	\draw (10, 2) node[above]{2};
	\draw (10, -2) node[below]{1};
	\draw[->, line width = 2] (12,-2)--(12,-1);
	\draw[line width = 2] (12,-2)--(12,0);
	\draw[->, line width = 2] (12,0)--(12,1);
	\draw[line width = 2] (12,0)--(12,2);
	\draw (12, 2) node[above]{4};
	\draw (12, -2) node[below]{3};
	
	\draw[decorate,decoration={snake, segment length=5, amplitude=1}](10,0)--(12,0);

	\draw (13, 0) node{$+$};
	
	\draw[->,dashdotted, line width = 1.5] (14,-2)--(14,-1);
	\draw[dashdotted, line width = 1.5] (14,-2)--(14,0);
	\draw[->, line width = 2] (14,0)--(14,1);
	\draw[ line width = 2] (14,0)--(14,2);
	\draw (14, 2) node[above]{4};
	\draw (14, -2) node[below]{1};
	\draw[->, line width = 2] (16,-2)--(16,-1);
	\draw[line width = 2] (16,-2)--(16,0);
	\draw[->, line width = 2] (16,0)--(16,1);
	\draw[line width = 2] (16,0)--(16,2);
	\draw (16, 2) node[above]{2};
	\draw (16, -2) node[below]{3};
	
	\draw[->,dashdotted, line width = 1.5] (22,-1)--(22,0.5);
	\draw[dashdotted, line width = 1.5] (22,-1)--(22,2);
	\draw [decorate, decoration = {brace,mirror}] (21,-1.5) --  (23,-1.5);
	\draw (22, -2) node[below]{$\mathcal{G}^{(1)}_{\rm HF}$};
	
	\draw[decorate,decoration={snake, segment length=5, amplitude=1}](14,0)--(16,0);
\end{tikzpicture}
%\caption{First order approximation for $\mathcal{G}^{(2,\rm T)}$}
	\label{fig:G2Tfirst}
\end{equation}

As the two-body Green function is a four point function, one has four possible points over which one can take the time derivative, and hence four variants for the second hierarchy equation. The present truncation procedure of the hierarchy, where one discards $\mathcal{G}^{(3,\rm T)}$ and $\mathcal{G}^{(1)}\cdot \mathcal{G}^{(2,\rm T)}$ terms excepted HF type terms, can then be derived in four different ways. The chosen point is visible in the diagrams because it is connected to the rest of the diagram by a HF dash-dotted propagator while the others are connected by full propagators. Of course, considering the first hierarchy equation, with any of these 4 possible $\mathcal{G}^{(2,\rm T)}$, one gets a self-consistent equation for $\mathcal{G}^{(1)}$. At the critical point (if it exists), one can check that the $1/r$-decay of the off-diagonal density matrix observed both in the ideal and HF cases is slightly modified by a multiplicative logarithmic term. 

\subsection{Symmetry-preserving closure and $n$-body correlations at all orders}

A severe drawback of the approximation~\eqref{fig:G2Tfirst} for $\mathcal{G}^{(3,\rm T)}$ is the breaking of the symmetry of $\mathcal{G}^{(2,\rm T)}$ with respect to the exchange of the points $1$ and~$3$. In order to cure this spurious asymmetry, it is crucial to keep a non-vanishing $\mathcal{G}^{(3,\rm T)}$ in the second hierarchy equation. In order to determine what form of $\mathcal{G}^{(3,\rm T)}$ restores the symmetry for $\mathcal{G}^{(2)}$, it is convenient to recast the hierarchy equations in the general form
\be
\label{eq:defO}
\hat{\mathcal{O}}_{1} f = g
\ee
with $\hat{\mathcal{O}}_{1} = \hbar \frac{\partial}{\partial \tau_{1}} - \frac{\hbar^2}{2m}\Delta_{1} - \mu$.
Then one uses the standard expression of $f$ as a convolution of the source term $g$ with the Green function of $\hat{\mathcal{O}}_{1}$, which is nothing but $\mathcal{G}_{0}$. As shown in Appendix \ref{AA}, this leads to the required form of $\mathcal{G}^{(3,\rm T)}$ that provides a symmetric $\mathcal{G}^{(2,\rm T)}$.
We obtain
\begin{equation}
\begin{tikzpicture}[scale=0.5,baseline={([yshift=-.5ex]current bounding box.center)},vertex/.style={anchor=base}]
	\draw[] (6,-2) to[bend left=-20] (6,2);
	\draw[] (8,-2) to[bend left=20] (8,2);
	\draw[] (6,-2)  to[bend left=40] (8,-2);
	\draw[] (6,2)  to[bend left=-40] (8,2);
	\draw (6, 2) node[right]{$\bullet$};
	\draw (8, 2) node[left]{$\bullet$};
	\draw (6, 2) node[left]{2};
	\draw (8, 2) node[right]{4};
	\draw (6, -2) node[left]{1};
	\draw (8, -2) node[right]{3};
	\draw (7,0) node{T};

	\draw (9, 0) node{$=$};
	
	\draw[->,line width = 2] (10,-2)--(10,-1);
	\draw[line width = 2] (10,-2)--(10,0);
	\draw[->, line width = 2] (10,0)--(10,1);
	\draw[ line width = 2] (10,0)--(10,2);
	\draw (10, 2) node[above]{2};
	\draw (10, -2) node[below]{1};
	\draw[->, line width = 2] (12,-2)--(12,-1);
	\draw[line width = 2] (12,-2)--(12,0);
	\draw[->, line width = 2] (12,0)--(12,1);
	\draw[line width = 2] (12,0)--(12,2);
	\draw (12, 2) node[above]{4};
	\draw (12, -2) node[below]{3};
	
	\draw[decorate,decoration={snake, segment length=5, amplitude=1}](10,0)--(12,0);

	\draw (13, 0) node{$+$};
	
	\draw[->,line width = 2] (14,-2)--(14,-1);
	\draw[line width = 2] (14,-2)--(14,0);
	\draw[->, line width = 2] (14,0)--(14,1);
	\draw[ line width = 2] (14,0)--(14,2);
	\draw (14, 2) node[above]{4};
	\draw (14, -2) node[below]{1};
	\draw[->, line width = 2] (16,-2)--(16,-1);
	\draw[line width = 2] (16,-2)--(16,0);
	\draw[->, line width = 2] (16,0)--(16,1);
	\draw[line width = 2] (16,0)--(16,2);
	\draw (16, 2) node[above]{2};
	\draw (16, -2) node[below]{3};
	
	\draw[decorate,decoration={snake, segment length=5, amplitude=1}](14,0)--(16,0);
\end{tikzpicture}
%\caption{Symmetric first order approximation for $\mathcal{G}^{(2,\rm T)}$}
	\label{fig:G2Tholz}
\end{equation}
which is analogous to eq.~\eqref{fig:G2Tfirst}, but where the HF propagator is replaced by the full Green function $\mathcal{G}^{(1)}$.
This simple form of $\mathcal{G}^{(2,\rm T)}$ can alternatively be derived by using two other formalisms which do not rely on the hierarchy equation : 
\begin{enumerate}[label=(\roman*)]
\item the Feynman diagram expansion representation of $\mathcal{G}^{(2,\rm T)}$
\item the classical polymer gas equivalent to the quantum system \cite{MP2003,MP2005} where the $n$-body Green function are represented by $n$ opened filaments (impurities) immersed in a gas of loops.
\end{enumerate}

\subsection{Proper self-energy within the hierarchy approach}

It is interesting to derive the proper self-energy associated with any approximation for $\mathcal{G}^{(2)}$. We recast the first hierarchy equation into the form \eqref{eq:defO}, i.e.
\begin{align}
\hat{\mathcal{O}}_{1} \mathcal{G}^{(1)}(1 \mid 2) = \delta(\bm{r_{1}}-\bm{r_{2}}) \delta(\tau_{1}-\tau_{2})  - \int d\bm{r_{3}}  V(\abs{\bm{r_{3}}-\bm{r_{1}}}) \mathcal{G}^{(2)}(1, 3_{1} \mid 3_{1^+},2) \label{eq:firsteqfull}
\end{align}
As any equation of the form \eqref{eq:defO}, the solution of (\ref{eq:firsteqfull}) reduces to a convolution of the source term with the Green function of $\hat{\mathcal{O}}_{1}$ which is nothing but $\mathcal{G}^{(1)}_{0}$. This provides
\begin{align}
\mathcal{G}^{(1)}(1 \mid 2) = \mathcal{G}^{(1)}_{0}(1 \mid 2) - \int d\bm{r_{3}} \int d\bm{r_{5}} V(\abs{\bm{r_{3}}-\bm{r_{5}}}) \mathcal{G}^{(2)}(5_{1}, 3_{1} \mid 3_{1^+},2) \mathcal{G}^{(1)}_{0}(1 \mid 5_{1}) \label{eq:firstsolexact}
\end{align}
Using \eqref{eq:decompo} and setting $\mathcal{G}^{(2,\rm T)}=0$ into (\ref{eq:firstsolexact}), the resulting equation becomes identical to the Dyson equation (\ref{eq:Dyson}) with the HF proper self-energy \eqref{Fig:HF_proper}, as it should be.

The exact solution \eqref{eq:firstsolexact} can be represented diagrammatically as
\begin{equation}
\begin{tikzpicture}[scale=0.5,baseline={([yshift=-.5ex]current bounding box.center)},vertex/.style={anchor=base}]

	\draw[->, line width = 2] (14,-2)--(14,0);
	\draw[line width = 2] (14,-2)--(14,2);
	\draw (14, 2) node[above]{2};
	\draw (14, -2) node[below]{1};
	
	\draw (16,0) node{$=$};

	\draw[->] (18,-2)--(18,0);
	\draw[] (18,-2)--(18,2);
	\draw (18, 2) node[above]{2};
	\draw (18, -2) node[below]{1};
	
	\draw (20,0) node{$-$};
	
	\draw[] (22,-2) to[out=-30,in=0](22,-3);
	\draw[] (22,-2) to[out=-120,in=-180](22,-3);
	\draw[] (24,-2) to[bend left=10] (24,2);
	\draw[] (22,-2)  to[bend left=90] (24,-2);
	\draw[] (22,-2)  to[out=150,in=-100] (24,2);
	\draw (22, -2.5) node{$\bullet$};
	\draw (24, 2) node[left]{$\bullet$};
	\draw (24, 2) node[right]{2};
	\draw (21.5, -2) node[left]{3};
	\draw (24, -2) node[right]{5};
	\draw (24, -4) node[right]{1};
	\draw[decorate,decoration={snake, segment length=5, amplitude=1}](22,-2)--(24,-2);
	\draw[->] (24,-4)--(24,-3);
	\draw[] (24,-4)--(24,-2);
\end{tikzpicture}
%\caption{Exact solution of the first equation of the hierarchy}
\label{fig:exactfirst}
\end{equation}
Since $\mathcal{G}^{(2)}$ is evaluated at two almost identical arguments ($3_1$ and $3_{1^+}$), one has brought together two corners of the sheet that represents $\mathcal{G}^{(2)}$ as depicted here:
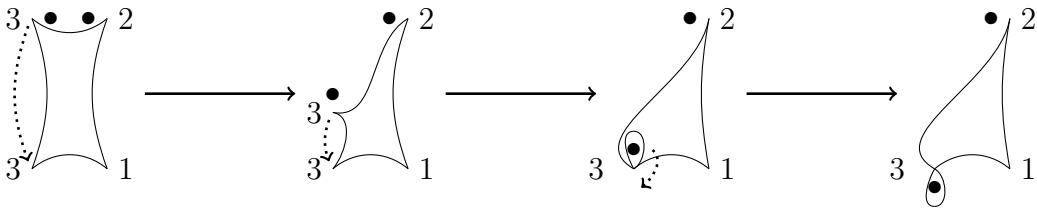
\begin{center}
\begin{tikzpicture}[scale=0.5]
	\draw[] (6,-2) to[bend left=-20] (6,2);
	\draw[] (8,-2) to[bend left=20] (8,2);
	\draw[] (6,-2)  to[bend left=40] (8,-2);
	\draw[] (6,2)  to[bend left=-40] (8,2);
	\draw (6, 2) node[right]{$\bullet$};
	\draw (8, 2) node[left]{$\bullet$};
	\draw (6, 2) node[left]{3};
	\draw (8, 2) node[right]{2};
	\draw (6, -2) node[left]{3};
	\draw (8, -2) node[right]{1};
	
	\draw[->, line width = 1, dotted] (5.9,1.8) to[bend right=20] (5.9,-1.8);
	
	\draw[->, line width = 1] (9,0)--(13,0);
	
	\draw[] (14,-2) to[out=50,in=-10] (14,-0.5);
	\draw[] (16,-2) to[bend left=20] (16,2);
	\draw[] (14,-2)  to[bend left=40] (16,-2);
	\draw[] (14,-0.5)  to[out=-10,in=-150] (16,2);
	\draw (14, -0.5) node[above]{$\bullet$};
	\draw (16, 2) node[left]{$\bullet$};
	\draw (14, -0.5) node[left]{3};
	\draw (16, 2) node[right]{2};
	\draw (14, -2) node[left]{3};
	\draw (16, -2) node[right]{1};
	
	\draw[->, line width = 1, dotted] (13.9,-0.7) to[bend right=20] (13.9,-1.8);
	
	\draw[->, line width = 1] (17,0)--(21,0);
	
	\draw[] (22,-2) to[out=30,in=0](22,-1);
	\draw[] (22,-2) to[out=120,in=-180](22,-1);
	\draw[] (24,-2) to[bend left=10] (24,2);
	\draw[] (22,-2)  to[bend left=40] (24,-2);
	\draw[] (22,-2)  to[out=150,in=-100] (24,2);
	\draw (22, -1.5) node{$\bullet$};
	\draw (24, 2) node[left]{$\bullet$};
	\draw (24, 2) node[right]{2};
	\draw (21.5, -2) node[left]{3};
	\draw (24, -2) node[right]{1};
	
	\draw[->, line width = 1, dotted] (22.5,-1.5) to[out=-40,in=40]  (22.2,-2.5);
	
	\draw[->, line width = 1] (25,0)--(29,0);
	
	\draw[] (30,-2) to[out=-30,in=0](30,-3);
	\draw[] (30,-2) to[out=-120,in=-180](30,-3);
	\draw[] (32,-2) to[bend left=10] (32,2);
	\draw[] (30,-2)  to[bend left=40] (32,-2);
	\draw[] (30,-2)  to[out=150,in=-100] (32,2);
	\draw (30, -2.5) node{$\bullet$};
	\draw (32, 2) node[left]{$\bullet$};
	\draw (32, 2) node[right]{2};
	\draw (29.5, -2) node[left]{3};
	\draw (32, -2) node[right]{1};
\end{tikzpicture}
\end{center}
(In the last step, one moves upwards the pinched corner 3 and downwards the loop-shaped boundary of the sheet). The loop enclosing a dot identifies the point of $\mathcal{G}^{(2)}$ that is both an input and an output point.

Inserting now the approximation \eqref{fig:G2Tholz} for $\mathcal{G}^{(2,\rm T)}$ in the exact solution \eqref{fig:exactfirst} of the first hierarchy equation~\cite{MT2006}, we again retrieve the Dyson equation with a specific proper self-energy, which includes the HF term \eqref{Fig:HF_proper} and the following two terms
\begin{equation}
\begin{tikzpicture}[scale=0.6,baseline={([yshift=-.5ex]current bounding box.center)},vertex/.style={anchor=base}]
	\draw (2.5,0.8) node[left]{$\Sigma^\star_{(2,\rm T)}$};
	
	\draw (3,1) node{$=$};

	\draw[decorate,decoration={snake, segment length=5, amplitude=1}](4,-1)--(5,-1);
	\draw[decorate,decoration={snake, segment length=5, amplitude=1}](4,3)--(5,3);
	\draw[->, line width = 2] (6.01,0.9)--(6.01,1);
	\draw[->, line width = 2] (4.41,1)--(4.41,0.9);
	\draw[ line width = 2] (5,-1) to[out=30,in=-30](5,3);
	\draw[ line width = 2] (5,3) to[out=-120,in=120](5,-1);
	\draw[ line width = 2] (4,-1)--(4,3);
	\draw[->, line width = 2] (4,-1)--(4,1);

	\draw (6.5,1) node{$+$};

	\draw[->, line width = 2] (8,-0.2)--(8,-0.2);
	\draw[->, line width = 2] (8,0.6)--(8,1);

	\draw[->, line width = 2] (8,2.2)--(8,2.6);
	\draw[ line width = 2] (8,-1)--(8,3);

	\draw[decorate, decoration={snake, segment length=5, amplitude=1}] (8,0) to[bend left=-90] (8,3);
	\draw[decorate, decoration={snake, segment length=5, amplitude=1}] (8,2) to[bend left=-90] (8,-1);
\end{tikzpicture}
%\caption{Baym et al. proper self-energy correction to the HF approximation}
\label{fig:properbaym}
\end{equation}
These two corrections are precisely the proper self-energy first introduced and estimated by Baym et al. \cite{BBH2001}. As Baym et al. have shown in the case of a dilute gas, this approximation leads to a modification of the exponent $s$ in the large-distance critical behavior, which becomes $s=3/2$ instead of $1$ in the ideal and HF cases. Moreover, at a given fixed low-density, the critical temperature deviates from its ideal value by a small shift.

\section{Concluding comments and perspectives}

Throughout this paper, we have analyzed the correspondence between the standard Feynman diagrammatic expansion and the hierarchy equations for the imaginary-time Green functions. Such correspondence has been illustrated within successive approximations, which can be equivalently formulated in both formalisms.

Closing a hierarchy of equations consists in truncating it at some level by introducing a sensible approximation. While the Hartree-Fock approximation amounts to neglecting entirely two-body correlations ($\mathcal{G}^{(2,\rm T)} = 0$), we have found that adequate closures beyond HF, i.e. closures that preserve the symmetry of $\mathcal{G}^{(2,\rm T)}$ with respect to the exchange of the points, require taking into account $n$-particle correlations at all orders $n$. We have made explicit the correspondance between the symmetry-preserving closure of the hierarchy and summing the classes of diagrams considered in Ref.~\cite{BBH2001}. 

The hierarchy approach enriches the usual Feynman diagram picture by providing a clear more global picture for the considered approximations. In particular, the ansatz introduced by Baym et al. \cite{BBH2001} for going beyond HF is equivalent to the natural simplest approximation for the truncated two-body Green-function $\mathcal{G}^{(2,\rm T)}$. This simple form of $\mathcal{G}^{(2,\rm T)}$ can be interpreted diagrammatically as resulting from two complete propagator $\mathcal{G}^{(1)}$ connected by a single interaction line (see eq.\eqref{fig:G2Tholz}).

The form~\eqref{fig:G2Tholz} can be generalized by considering diagrams with two $\mathcal{G}^{(1)}$'s and $n$ interaction lines, arranged with specific topological prescriptions (see forthcoming paper \cite{DBA1}). The corresponding representation of $\mathcal{G}^{(2, \rm T)}$ is a perturbation expansion with respect to the interaction potential where the reference ingredients are the complete Green function $\mathcal{G}^{(1)}$. Hence, the simple form~\eqref{fig:G2Tholz} turns to be the lowest-order contribution in this representation, keeping in mind that $\mathcal{G}^{(1)}$ itself incorporates contributions of arbitrary orders in the interaction.

Since the contributions of short-range interactions can reasonably be expected to provide small contributions at low density, the insertion of the simple $\mathcal{G}^{(2, \rm T)}$ [eq.~\eqref{fig:G2Tholz}] into the first hierarchy equation for $\mathcal{G}^{(1)}$ should provide the next corrections to the HF approximation, as argued by Baym et al. within the proper self-energy approach. In a forthcoming paper \cite{DBA2}, we will carefully investigate the corresponding non-linear integro-differential equation for $\mathcal{G}^{(1)}$, both on the numerical and analytical sides. This will complement the results of Baym et al. : in particular the contribution of the dynamical non-zero Matsubara frequencies, remain to be properly determined, while the ultraviolet singularities induced by a $\delta$-potential have to be treated explicitly.

The diagrammatic representation of $\mathcal{G}^{(2, \rm T)}$ in terms of $\mathcal{G}^{(1)}$ and of the interaction \cite{DBA1} paves the way for improving the HF theory to higher order in the interaction, and ultimately in the density. For short range potentials, it would be quite interesting to determine the critical exponent $s_{p}$ obtained for each successive approximation of order $p$ in the interaction (limit $s_{p}$ when $p \to \infty$ ?). For long-range potentials with the Kac form, {\it i.e.}in the limit $\gamma \to 0$, the very existence of the critical point is questionnable since HF theory predicts the breakdown of the BE condensation \cite{APS2019}.

\appendix

\section{}
\la{AA}

The exact solution of the second equation (\ref{eq:second_eq}) of the hierarchy, which involves $\mathcal{G}^{(3)}$, can be represented as
\begin{equation}
\begin{tikzpicture}[scale=0.5,baseline={([yshift=-.5ex]current bounding box.center)},vertex/.style={anchor=base}]

		\draw[] (6,-2) to[bend left=-20] (6,2);
	\draw[] (8,-2) to[bend left=20] (8,2);
	\draw[] (6,-2)  to[bend left=40] (8,-2);
	\draw[] (6,2)  to[bend left=-40] (8,2);
	\draw (6, 2) node[right]{$\bullet$};
	\draw (8, 2) node[left]{$\bullet$};
	\draw (6, 2) node[left]{2};
	\draw (8, 2) node[right]{4};
	\draw (6, -2) node[left]{1};
	\draw (8, -2) node[right]{3};
	
	\draw (9,0) node{$=$};
	
	\draw[->] (10,-2)--(10,0);
	\draw[] (10,-2)--(10,2);
	\draw (10, 2) node[above]{2};
	\draw (10, -2) node[below]{1};
	
	\draw[->, line width = 2] (12,-2)--(12,0);
	\draw[line width = 2] (12,-2)--(12,2);
	\draw (12, 2) node[above]{4};
	\draw (12, -2) node[below]{3};
	
	\draw (14,0) node{$+$};
	
	\draw[->] (16,-2)--(16,0);
	\draw[] (16,-2)--(16,2);
	\draw (16, 2) node[above]{4};
	\draw (16, -2) node[below]{1};

	\draw[->,line width = 2] (18,-2)--(18,0);
	\draw[line width = 2] (18,-2)--(18,2);
	\draw (18, 2) node[above]{2};
	\draw (18, -2) node[below]{3};
	
	\draw (20,0) node{$-$};
	
	\draw[] (22,-2) to[out=-30,in=0](22,-3);
	\draw[] (22,-2) to[out=-120,in=-180](22,-3);
	\draw[] (24,2) to[bend left=-90] (26,2);
	\draw[] (22,-2)  to[bend left=90] (24,-2);
	\draw[] (22,-2)  to[out=150,in=-100] (24,2);
	\draw[] (26,-2) to[bend left=10] (26,2);
	\draw[] (24,-2)  to[bend left=90] (26,-2);
	\draw (22, -2.5) node{$\bullet$};
	\draw (24, 2) node[left]{$\bullet$};
	\draw (26, 2) node[left]{$\bullet$};
	\draw (24, 2) node[right]{2};
	\draw (21.5, -2) node[left]{5};
	\draw (24, -2) node[right]{6};
	\draw (24, -4) node[right]{1};
	\draw (26, 2) node[right]{4};
	\draw (26, -2) node[right]{3};
	\draw[decorate,decoration={snake, segment length=5, amplitude=1}](22,-2)--(24,-2);
	\draw[->] (24,-4)--(24,-3);
	\draw[] (24,-4)--(24,-2);
\end{tikzpicture}
%\caption{Exact solution of the second equation of the hierarchy}
\label{fig:exactsecond}
\end{equation}
similarly to eq.~\eqref{fig:exactfirst}~\cite{MT2006}.
 Expanding $\mathcal{G}^{(3)}$ in terms of $\mathcal{G}^{(1)}$, $\mathcal{G}^{(2,\rm T)}$, and $\mathcal{G}^{(3,\rm T)}$, one can easily make appear the asymmetric approximation~\eqref{fig:G2Tfirst} and also the terms needed to replace the free propagators $\mathcal{G}^{(1)}_{0}(1\vert2)$ and $\mathcal{G}^{(1)}_{0}(1\vert4)$ by full propagators. To retrieve the symmetric approximation one needs to replace an HF propagator with a full propagator. 
 \begin{figure}[h!]
 \centering
 \begin{tikzpicture}[scale=0.5]
	 \draw[] (22,-2) to[out=-30,in=0](22,-3);
	\draw[] (22,-2) to[out=-120,in=-180](22,-3);
	\draw[] (24,-2) to[bend left=10] (24,2);
	\draw[] (22,-2)  to[bend left=90] (24,-2);
	\draw[] (22,-2)  to[out=150,in=-100] (24,2);
	\draw (22, -2.5) node{$\bullet$};
	\draw (24, 2) node[left]{$\bullet$};
	\draw[decorate,decoration={snake, segment length=5, amplitude=1}](22,-2)--(24,-2);
	\draw[->] (24,-4)--(24,-3);
	\draw[] (24,-4)--(24,-2);
	
	\draw[decorate,decoration={snake, segment length=5, amplitude=1}](24,2)--(26,2);
	
	\draw[->,line width = 2] (26,-2)--(26,0);
	\draw[line width = 2] (26,-2)--(26,2);
	
	\draw[->,line width = 2] (26,2)--(26,3);
	\draw[line width = 2] (26,2)--(26,4);
	
	\draw[->,line width = 2] (24,2)--(24,3);
	\draw[line width = 2] (24,2)--(24,4);
	
	\draw (24, -5) node{(a)};

	\draw[] (34,-2) to[out=-30,in=0](34,-3);
	\draw[] (34,-2) to[out=-120,in=-180](34,-3);
	\draw[] (36,4) to[bend left=-90] (38,4);
	\draw[] (34,-2)  to[bend left=90] (36,-2);
	\draw[] (34,-2)  to[out=150,in=-100] (36,4);
	\draw[] (38,-2) to[bend left=10] (38,4);
	\draw[] (36,-2)  to[bend left=90] (38,-2);
	\draw (34, -2.5) node{$\bullet$};
	\draw (36, 4) node[left]{$\bullet$};
	\draw (38, 4) node[left]{$\bullet$};
	\draw[decorate,decoration={snake, segment length=5, amplitude=1}](34,-2)--(36,-2);
	\draw[->] (36,-4)--(36,-3);
	\draw[] (36,-4)--(36,-2);
	
	\draw (36, -5) node{(b)};
	
\end{tikzpicture}
\caption{(a) Required term in terms of $\mathcal{G}^{(1)}$ and $\mathcal{G}^{(2)}$   (b) Required term in terms of $\mathcal{G}^{(1)}$ and $\mathcal{G}^{(3)}$ }
\label{fig:G3T}
\end{figure}
Using the exact solution~\eqref{fig:exactfirst} of the first equation of hierarchy, one looks for, in the $\mathcal{G}^{(3)}$-term, a diagram with the form shown in Figure \ref{fig:G3T} (a). Because this diagram contains two explicit interaction lines, it is necessarily part of $\mathcal{G}^{(3,\rm T)}$. As $\mathcal{G}^{(3,\rm T)}$ is evaluated in the exact solution~\eqref{fig:exactsecond} at the specific configuration represented in Figure \ref{fig:G3T} (b), one can determine the form that the three-body Green function must satisfy, namely
\begin{equation}
\notag
\begin{tikzpicture}[scale=0.7,baseline={([yshift=-.5ex]current bounding box.center)},vertex/.style={anchor=base}]
	\draw[] (10,-2) to[bend left=-20] (10,2);
	\draw[] (14,-2) to[bend left=20] (14,2);
	\draw[] (10,-2)  to[bend left=40] (12,-2);
	\draw[] (12,-2)  to[bend left=40] (14,-2);
	\draw[] (10,2)  to[bend left=-40] (12,2);
	\draw[] (12,2)  to[bend left=-40] (14,2);
	\draw (10, 2) node[right]{$\bullet$};
	\draw (14, 2) node[left]{$\bullet$};
	\draw (12, 2) node[left]{$\bullet$};
	\draw (12, 2) node[above]{2};
	\draw (10, 2) node[left]{6};
	\draw (14, 2) node[right]{4};
	\draw (12, -2) node[below]{1};
	\draw (10, -2) node[below]{5};
	\draw (14, -2) node[right]{3};
	
	\draw (12, 0) node{T};
	
	\draw (16,0) node{$=$};

	\draw[] (20,-2) to[bend left=10] (20,0);
	\draw[] (18,-2)  to[bend left=40] (20,-2);
	\draw[] (18,-2)  to[bend left = -20] (18,2);
	\draw (18, 2) node[right]{$\bullet$};
	\draw (20, 0) node[below right]{$\bullet$};
	
	\draw (19, -0.8) node{T};

	\draw[] (18,2)  to[out=-70,in=-130] (20,0);
	
	\draw[decorate,decoration={snake, segment length=5, amplitude=1}](20,0)--(22,0);
	
	\draw[->,line width = 2] (22,-2)--(22,-1);
	\draw[line width = 2] (22,-2)--(22,0);
	
	\draw[->,line width = 2] (22,0)--(22,1);
	\draw[line width = 2] (22,0)--(22,2);
	
	\draw[->,line width = 2] (20,0)--(20,1);
	\draw[line width = 2] (20,0)--(20,2);
	
	\draw (20, 2) node[above]{2};
	\draw (18, 2) node[left]{6};
	\draw (22, 2) node[right]{4};
	\draw (20, -2) node[below]{1};
	\draw (18, -2) node[below]{5};
	\draw (22, -2) node[right]{3};
	
  \draw (24,0) node{$+$};
 
	\draw[] (28,-2) to[bend left=10] (28,0);
	\draw[] (26,-2)  to[bend left=40] (28,-2);
	\draw[] (26,-2)  to[bend left = -20] (26,2);
	\draw (26, 2) node[right]{$\bullet$};
	\draw (28, 0) node[below right]{$\bullet$};
	
	\draw (27, -0.8) node{T};

	\draw[] (26,2)  to[out=-70,in=-130] (28,0);
	
	\draw[decorate,decoration={snake, segment length=5, amplitude=1}](28,0)--(30,0);
	
	\draw[->,line width = 2] (30,-2)--(30,-1);
	\draw[line width = 2] (30,-2)--(30,0);
	
	\draw[->,line width = 2] (30,0)--(30,1);
	\draw[line width = 2] (30,0)--(30,2);
	
	\draw[->,line width = 2] (28,0)--(28,1);
	\draw[line width = 2] (28,0)--(28,2);
	
	\draw (28, 2) node[above]{4};
	\draw (26, 2) node[left]{6};
	\draw (30, 2) node[right]{2};
	\draw (28, -2) node[below]{1};
	\draw (26, -2) node[below]{5};
	\draw (30, -2) node[right]{3};
\end{tikzpicture}
\end{equation}
\begin{equation}
\begin{tikzpicture}[scale=0.47]
	
	\draw (16,0) node{$=$};
 	
	\draw[decorate,decoration={snake, segment length=5, amplitude=1}](20,1)--(22,1);
	\draw[decorate,decoration={snake, segment length=5, amplitude=1}](18,-1)--(20,-1);
	
	\draw[->,line width = 2] (22,-3)--(22,-1);
	\draw[line width = 2] (22,-3)--(22,1);
	
	\draw[->,line width = 2] (22,1)--(22,2);
	\draw[line width = 2] (22,1)--(22,3);
	
	\draw[->,line width = 2] (20,1)--(20,2);
	\draw[line width = 2] (20,1)--(20,3);
	
	\draw[->,line width = 2] (20,-3)--(20,-2);
	\draw[line width = 2] (20,-3)--(20,-1);
	
	\draw[->,line width = 2] (20,-1)--(20,0);
	\draw[line width = 2] (20,-1)--(20,1);
	
	\draw[->,line width = 2] (18,-3)--(18,-2);
	\draw[line width = 2] (18,-3)--(18,-1);
	
	\draw[->,line width = 2] (18,-1)--(18,1);
	\draw[line width = 2] (18,-1)--(18,3);

	\draw (20, 3) node[left]{2};
	\draw (18, 3) node[left]{6};
	\draw (22, 3) node[right]{4};
	\draw (20, -3) node[left]{1};
	\draw (18, -3) node[left]{5};
	\draw (22, -3) node[right]{3};
	
  \draw (24,0) node{$+$};
 
	\draw[decorate,decoration={snake, segment length=5, amplitude=1}](28,1)--(30,1);
	\draw[decorate,decoration={snake, segment length=5, amplitude=1}](26,-1)--(28,-1);
	
	\draw[->,line width = 2] (30,-3)--(30,-1);
	\draw[line width = 2] (30,-3)--(30,1);
	
	\draw[->,line width = 2] (30,1)--(30,2);
	\draw[line width = 2] (30,1)--(30,3);
	
	\draw[->,line width = 2] (28,1)--(28,2);
	\draw[line width = 2] (28,1)--(28,3);
	
	\draw[->,line width = 2] (28,-3)--(28,-2);
	\draw[line width = 2] (28,-3)--(28,-1);
	
	\draw[->,line width = 2] (28,-1)--(28,0);
	\draw[line width = 2] (28,-1)--(28,1);
	
	\draw[->,line width = 2] (26,-3)--(26,-2);
	\draw[line width = 2] (26,-3)--(26,-1);
	
	\draw[->,line width = 2] (26,-1)--(26,1);
	\draw[line width = 2] (26,-1)--(26,3);

	\draw (28, 3) node[left]{2};
	\draw (26, 3) node[left]{6};
	\draw (30, 3) node[right]{4};
	\draw (28, -3) node[left]{5};
	\draw (26, -3) node[left]{1};
	\draw (30, -3) node[right]{3};
	
	 \draw (32,0) node{$+$};
 	
	\draw[decorate,decoration={snake, segment length=5, amplitude=1}](36,1)--(38,1);
	\draw[decorate,decoration={snake, segment length=5, amplitude=1}](34,-1)--(36,-1);
	
	\draw[->,line width = 2] (38,-3)--(38,-1);
	\draw[line width = 2] (38,-3)--(38,1);
	
	\draw[->,line width = 2] (38,1)--(38,2);
	\draw[line width = 2] (38,1)--(38,3);
	
	\draw[->,line width = 2] (36,1)--(36,2);
	\draw[line width = 2] (36,1)--(36,3);
	
	\draw[->,line width = 2] (36,-3)--(36,-2);
	\draw[line width = 2] (36,-3)--(36,-1);
	
	\draw[->,line width = 2] (36,-1)--(36,0);
	\draw[line width = 2] (36,-1)--(36,1);
	
	\draw[->,line width = 2] (34,-3)--(34,-2);
	\draw[line width = 2] (34,-3)--(34,-1);
	
	\draw[->,line width = 2] (34,-1)--(34,1);
	\draw[line width = 2] (34,-1)--(34,3);

	\draw (36, 3) node[left]{4};
	\draw (34, 3) node[left]{6};
	\draw (38, 3) node[right]{2};
	\draw (36, -3) node[left]{1};
	\draw (34, -3) node[left]{5};
	\draw (38, -3) node[right]{3};
	
  \draw (40,0) node{$+$};
 
	\draw[decorate,decoration={snake, segment length=5, amplitude=1}](44,1)--(46,1);
	\draw[decorate,decoration={snake, segment length=5, amplitude=1}](42,-1)--(44,-1);
	
	\draw[->,line width = 2] (46,-3)--(46,-1);
	\draw[line width = 2] (46,-3)--(46,1);
	
	\draw[->,line width = 2] (46,1)--(46,2);
	\draw[line width = 2] (46,1)--(46,3);
	
	\draw[->,line width = 2] (44,1)--(44,2);
	\draw[line width = 2] (44,1)--(44,3);
	
	\draw[->,line width = 2] (44,-3)--(44,-2);
	\draw[line width = 2] (44,-3)--(44,-1);
	
	\draw[->,line width = 2] (44,-1)--(44,0);
	\draw[line width = 2] (44,-1)--(44,1);
	
	\draw[->,line width = 2] (42,-3)--(42,-2);
	\draw[line width = 2] (42,-3)--(42,-1);
	
	\draw[->,line width = 2] (42,-1)--(42,1);
	\draw[line width = 2] (42,-1)--(42,3);

	\draw (44, 3) node[left]{4};
	\draw (42, 3) node[left]{6};
	\draw (46, 3) node[right]{2};
	\draw (44, -3) node[left]{5};
	\draw (42, -3) node[left]{1};
	\draw (46, -3) node[right]{3};
		
	\end{tikzpicture}
%\caption{Required form of $\mathcal{G}^{(3,\rm T)}$
%for the symmetric first order interaction approximation in $\mathcal{G}^{(2,\rm T)}$}
\label{fig:G3Trequired}
\end{equation}
With the closure provided by the above form of $\mathcal{G}^{(3,\rm T)}$, one gets the symmetric approximation~\eqref{fig:G2Tholz} for $\mathcal{G}^{(2,\rm T)}$, which is first order in the interaction.

An a priori surprising thing is that reestablishing the symmetry of $\mathcal{G}^{(2,\rm T)}$ requires taking an asymmetric $\mathcal{G}^{(3,\rm T)}$. Indeed, the diagrams in eq.~\eqref{fig:G3Trequired} would need to be evaluated at 18 different configurations to be symmetric, whereas only 4 specific configurations intervene in~\eqref{fig:G3Trequired}. In fact, this is not a problem and one can symmetrize $\mathcal{G}^{(3,\rm T)}$ by including all 18 diagrams (configurations). Indeed, since the diagrams are evaluated in eq.~\eqref{fig:exactsecond} at the specific configuration of Fig.~1b, the 14 additional diagrams lead to a negligible third-order contribution to $\mathcal{G}^{(2,\rm T)}$.
In the same way, while the 6 diagrams with form $\mathcal{G}^{(1)} \cdot \mathcal{G}^{(1)} \cdot \mathcal{G}^{(1)}$ in $\mathcal{G}^{(3)}$ are used to reproduce the symmetric first order approximation for $\mathcal{G}^{(2,\rm T)}$, only 2 diagrams which come from $\mathcal{G}^{(1)} \cdot \mathcal{G}^{(2,\rm T)}$ are used while the 7 others contribute to $\mathcal{G}^{(2,\rm T)}$ at the second order in the interaction.
pp.
The symmetric approximation~\eqref{fig:G2Tholz} for $\mathcal{G}^{(2,\rm T)}$ implies the form~\eqref{fig:G3Trequired} for $\mathcal{G}^{(3,\rm T)}$ which is analogous of that of $\mathcal{G}^{(2,\rm T)}$. One can reproduce the same procedure on the third equation of the hierarchy and look for the required form of $\mathcal{G}^{(4,\rm T)}$ to obtain the required form of $\mathcal{G}^{(3,\rm T)}$, and so on. $n$-particle correlations for all $n$ are therefore taken into account in the symmetric approximation~\eqref{fig:G2Tholz}. It is worth pointing out that any closure of the hierarchy is equivalent to summing some classes of Feynman diagrams (as shown in Section \ref{SctIVA}), but the converse is not true: a summation of classes of Feynman diagrams does not always correspond to some closure of the hierarchy.

\end{document}